\documentclass[11pt,twoside]{article}

\usepackage{asp2006}
\usepackage{epsf}
\usepackage{psfig}
\usepackage{lscape}

\begin{document}
\title{Protocols for Scholarly Communication}   
\author{Alberto Pepe and Joanne Yeomans}   
\affil{User Documentation Group and Scientific Information Service, CERN, 1211 Geneva 23, Switzerland}    

\begin{abstract}
CERN, the European Organization for Nuclear Research, has operated an
institutional preprint repository for more than 10 years. The
repository contains over 850,000 records of which more than 450,000
are full-text OA preprints, mostly in the field of particle physics,
and it is integrated with the library's holdings of books, conference
proceedings, journals and other grey literature. In order to encourage 
effective propagation and open access to scholarly material, CERN is 
implementing a range of innovative library services into its document 
repository: automatic keywording, reference extraction, collaborative 
management tools and bibliometric tools. Some of these services, such 
as user reviewing and automatic metadata extraction, could make up an 
interesting testbed for future publishing solutions and certainly 
provide an exciting environment for e-science possibilities. The 
future protocol for scientific communication should naturally guide 
authors towards OA publication and CERN wants to help reach a full 
open access publishing environment for the particle physics community 
and the related sciences in the next few years.
\end{abstract}

\section{Preamble}

\par CERN has been an active leader of particle physics research for 
just over fifty years. During that period, physicists have led and 
developed a preprint-sharing culture which in its present, electronic 
incarnation is capturing the imagination of open access (OA) enthusiasts, 
academic librarians and repository managers, research institution 
directors, and the researchers themselves in fields other than physics, 
who wish to speed up, and open up, the process of scholarly communication. 
We wish to share with the library community the ways in which the CERN 
Library and Document Server teams have successfully managed and filled 
their preprint repository, the directions in which the management of 
this repository is now moving, and the implications this has for the 
future of scholarly communication in the particle physics field. Using 
technical protocols to fill and manage the repository in turn changes 
the protocols governing the communication environment and the processes 
of writing, publishing and reading scientific documents become more 
integrated.

\section{Document management overview}

\par Each year CERN authors produce around 2000 original papers, and make 
about 10,000 conference contributions. In an extension of the original 
paper-era mandate, the library is charged with collecting evidence of this 
activity in its various forms; this includes not just the collection of 
electronic versions of the published papers, but also copies of slides, 
posters, lecture notes and other kinds of contributions. Collecting such 
items not only enables the library to make them available to the world-wide 
physics community but also makes it possible to archive the material to 
protect the history of CERN's existence.

\par In order to store and manage these documents, CERN created in 1993 
an institutional repository which now exists under the name of the CERN 
Document Server (CDS)\footnote{http://cdsweb.cern.ch/}, an instance of the CDS 
Invenio repository software\footnote{http://cdsware.cern.ch/invenio/}. 
The repository was merged with the Library's 
catalogue software in 1996 so that the collections are now displayed in 
a single interface. The total number of records in CDS is approaching 
one million and is growing by an average of 280 each day. With the small 
number of cataloging staff this number of records could not be created 
manually nor handled individually and so the preprint management team 
has devised a number of technical solutions for the delivery of catalogue 
records from other databases and institutional and subject repositories, 
and for the updating and enhancement of those records.  

\par To facilitate searching, the materials are subdivided into 
collections of physically similar items, and collections of interest to 
particular groups. So, a user can easily limit his or her search to find 
a book on the library shelves, or to find photos, meeting minutes, 
articles and conference reports on the design of the ATLAS detector.

\par CDS also has quick links to the submission interface whereby files 
can be uploaded to the server by any member of CERN staff, and a file 
conversion service which enables users to convert files into a format 
that is more easily archived such as the portable document format, or 
ASCII text.

\section{CERN Document Server facts and figures}

\par Although CERN authors are required by CERN rules to submit their 
documents to the Library \citep{pepecern01}, in reality this happens for 
less than half of the known publications. The rest are mostly detected 
by the performance of regular web searches from where the full-text is 
also retrieved when possible. Due to these efforts, the Library believes 
it locates almost 100\% of the metadata of CERN-authored documents, and 
in recent years has managed to also obtain OA full-text versions of 
over 70\% of those documents \citep{pepeyeom06}. Retro-scanning projects 
both at CERN and Japan's KEK library, and permission from APS and other 
publishers to allow download of published versions, have brought many 
thousands of older documents into the digital OA environment such that 
around 54\% of all CERN scientific documents published since 1954 are 
now available for free in full-text versions.

\par In total there are over 850,000 bibliographic records including 
records for library books and journals, of which 450,000 have electronic 
full-text, OA documents attached. The metadata for virtually all CDS 
records is available free to those wishing to harvest and use it. The 
only exceptions are records of sensitive and financial nature and managerial 
documents. 

\par As CERN already has a mandate for self-archiving, other methods 
are being planned and implemented to both directly and indirectly 
encourage submission. Direct emailing of authors is an option for 
chasing missing items, and a promotion campaign has been directed at 
raising awareness of OA issues and reminding about submission 
procedures. These are both aimed at directly improving author 
submission. However, more subtle techniques are also used: the 
development of the repository in such a way that it becomes 
indispensable to authors in the process of their work and authoring 
duties which it is hoped will indirectly persuade them of the benefits 
of placing all their materials in a single location. The repository is 
now extremely well-used by CERN scientists to the extent that it 
receives 20,000 unique visitors each month (staff on site number around 
6,000 at any one time) and over 200,000 searches. In order to maintain 
this demand it is important that the service is continually enhanced 
to integrate with new services and match the expectations created by new 
technical possibilities in the information environment. Such repository 
enhancements are explained further in this article. 

\section{Repository-centred library services}

\subsection{Metadata enhancement}
\par The deluge of digital information experienced by the digital
world in the past decade has created a need for transparent and
effective means of data organization and mining. Within the digital
library environment, metadata has become the core component for
document reposition and dissemination.  Accurate, intelligible and
rich metadata helps the pinpointing of library objects within an
archive as well as across federated search engines: it is found that
libraries favour hosting metadata-rich objects, thus improving the
long-term access and preservation via replication at multiple sites.

\par At CERN, the metadata format of choice is MARCXML, a flexible,
extensible format which perfectly accommodates the needs of describing
heterogeneous and complex library objects. In the case of manuscripts
and articles - the predominant object type in the CDS
- there exists a set of basic metadata tags (such as author, title,
year, etc.) that are essential in order to store the object in the
archive; these are usually directly input by the author upon
submission or are acquired through an OAI-compliant harvest and
subsequently checked by library cataloguers.  At this point of the
data acquisition process, certain tools are evoked which attempt to
transcend the basic set of metadata tags by automatically
extracting additional information from the document's
fulltext. At the moment these tools typically perform automatic 
keywording and reference extraction. Although a detailed account of 
these procedures is beyond the scope of this paper, an outline of 
their basic \textit{modus operandi} gives an idea of their relevance 
to the overall organisation and usefulness of the
repository. For more information see \cite{pepeholt06} and \cite{pepeclai01} 
respectively.

\subsubsection{Automatic keywording}
\par The automatic keyword extraction aims at producing a set of
keywords that describe the fundamental concepts of a document. The
enrichment of metadata with controlled, subject-specific terms aids
document cataloging and indexing. In this perspective, CERN is
developing in conjunction with the DESY\footnote{Deutsches Elektronen
Synchrotron, Hamburg, Germany} Library a taxonomy of High Energy
Physics to be used with an automatic classification system. The
taxonomy, expressed in SKOS\footnote{Simple Knowledge Organisation
System} syntax, a dialect of RDF and XML markup
languages, contains more than 2500 basic terms and is implemented with
more than 15000 keyword combinations and key-chains, to satisfy the
needs and the classification methods sought by this specific
scientific community. By using a powerful phrase-matching mechanism on
top of this semantically-rich, well-structured knowledge base, we
are able to perform accurate assignment of controlled indexes to the
documents in the archive based on keyword occurrence and similar
algorithms. A typical output obtained with such a system is shown in
Table ~1. The benefit provided by such metadata enhancement is
two-fold: first, the controlled terms automatically generated provide
extra cataloging information on top of author and library indexes - a
clear added value; second, the taxonomy allows for the specification
of relations and hierarchies, so that the metadata generated can be
used to automatically create clusters of similar documents and refine
search capabilities.

\begin{table}[!ht]
\label{keywords}
\caption{Automatic keyword extraction: sample output for arXiv:gr-qc/0607062 - \textit{Classical and Quantum Dilaton Gravity in Two Dimensions with Fermions}}
\smallskip
\begin{center}
{\small
\begin{tabular}{ll}
\tableline
\noalign{\smallskip}
Occurrence & Thesaurus keyword\\
\noalign{\smallskip}
\tableline
\noalign{\smallskip}
39  & magnetic moment\\
38 [73, 121]  & gravitation, dilaton \\
32  & effective action\\
28 [25, 41]  & quantization, nonperturbative \\
21  & ghost\\
20  & Poisson bracket\\
16 [12, 44] & field theory, scalar \\
15  & Minkowski\\
14  & bosonization\\
12 [65, 71] & fermion, Dirac \\
\noalign{\smallskip}
\tableline
\end{tabular}
}
\end{center}
\end{table}

\subsubsection{Reference extraction}
\par The reference extraction aims at automatically detecting the
bibliography list from a document's fulltext. The information
retrieved is then used to enhance the metadata and produce
bibliometric evaluations such as those described in section \ref{usage}  The
methodology used at CERN consists of three steps: 
\begin{enumerate}
\item the detection and extraction of the reference section from the full text
\item the recognition of single citation entries
\item the reformulation to standard and accurate citation format and 
thus the linking to the cited sources
\end{enumerate}
The first step involves some text-parsing methods to isolate the
portion of the document that contains bibliographic information. Once
this has been localised, the style and structure of the bibliography
is interpreted. This involves recognising every single citation entry
and for each one, reconstructing its bibliographic information such as
title, author, report number and Internet address. In order to improve
the quality and accuracy of the output, the reference extraction
operates on top of a knowledge base which contains alternative forms
of scientific journal names and report numbers. An excerpt of a sample
knowledge base is depicted in Table ~2. By using this
information, the system tries to match the author-created citation entry 
with a known on-line journal volume and thus uses the publisher's system 
of URL creation to automatically
generate a working URL, linking back to the cited source.

\begin{table}[!ht]
\label{kb}
\caption{Journal titles: excerpt from knowledge base}
\smallskip
\begin{center}
{\small
\begin{tabular}{ll}
\tableline
\noalign{\smallskip}
Journal title & Alternative forms\\
\noalign{\smallskip}
\tableline
\noalign{\smallskip}
Astron. Astrophys. & A \& A, A A, A A LETT, A A LETTERS, AAL\\
ACM Comput. Surv. & ACM COMPUTING SURVEYS\\
ACM SIGPLAN Not. & ACM SIGPLAN NOTICES, ACM SN\\
IEEE J. Quantum Electron. & IJQE\\
J. High Energy Phys. & JHEP\\
New Sci. & NEW SCIENTIST\\
Phys. Rev., A  & PHYSICAL REVIEW A, PHYS REV A, PRA\\
\noalign{\smallskip}
\tableline
\end{tabular}
}
\end{center}
\end{table}

\subsection{Collaborative management}
\par Co-operative tools are rapidly being introduced in many modern 
on-line content-management systems. The types of social tools currently 
offered on a large scale include services for social editing (wikis), 
basket management, open content evaluation and user interaction 
(forums and message-boards). The benefits of adopting 
a collaborative infrastructure on top of a digital library system are 
numerous, yet the essence of the service is unique: to provide a social 
environment in which users and communities interact with each other and 
actively participate in the management of digital content. 

\par In the field 
of high energy physics, the dissemination of scientific results among 
communities of researchers and academics has relied enormously in 
the past decade on the free circulation of electronic preprints and 
articles in specialised subject repositories, such as arXiv.org and 
SPIRES.  Although the importance and role of these large open archives 
remain invaluable, they still lack the dynamism offered by a 
user-centered, co-operative setting. The CERN Library and Document 
Server teams have therefore been increasingly interested in implementing 
a set of social tools into the institutional repository, to allow a 
higher degree of user interaction and active participation among the 
scientific community. Some of the services that are currently being 
deployed or are already in use are a) automatic user notification, 
b) basket and collection management and c) content rating and evaluation.

\subsubsection{Automatic user notification}
\par The automatic notification is a service intended for users that 
wish to be alerted whenever the repository is updated with certain 
documents. Notifications can enormously aid the localisation of 
specific types of material, as users can set up very highly defined alerts 
based on the output of particular searches. Such a system provides a 
useful service not only to the users who wish to receive updates, but 
also to the authors whose material becomes more visible to a specific 
intended audience.  The notifications are currently in the form of 
emails, sent upon request to the user, although an Atom/RSS feed 
facility will be available in the near future.  

\subsubsection{Basket management}
\par Basket and collection management is a groupware feature that 
was recently developed mainly to satisfy the business needs of companies 
in \textit{e}-commerce.  It has now rapidly evolved and become popular 
among many other user-oriented web services such as document, photo and 
music archives. This feature allows users and groups of users to collect 
digital library objects into organised baskets and collections and thus 
share their content with the broader community. As with user notification, 
this particular social feature improves the visibility of digital objects 
and facilitates their reuse and dissemination by allowing the export 
of baskets in bibliographic formats (Bib\TeX) and a variety of web feeds.  

\subsubsection{Content evaluation}
\par The rating and evaluation of digital library content is 
undoubtedly the most influential, yet controversial, social feature 
which will be offered to CDS users. The basic concept is to allow 
open discussion and review of all archived content. This paradigm, 
already extensively used in non-academic domains such as music, books 
and movie review services, is struggling to gain wide acceptance from 
the academic and scientific communities whose research evaluation 
channels have been historically bound to traditional peer reviewing 
methods, published letters and private communication. In the CDS prototype, all 
users are allowed to review a library object, rate it and comment on 
other user reviews, thus opening active discussion and interaction in a 
message-board fashion. In an attempt to reduce the misuse, all users can 
also report malicious submissions at any time --- to allow human 
interception.  The introduction of an open document review and rating 
scheme in the CDS repository is not intended to discrown the validity and 
efficiency of peer review, but rather provide an increased level of 
feedback and review in parallel to the traditional evaluation and commenting 
methods. Moreover, the formal presence of such a system may prove to be 
an incentive for more authors to deposit their preprints and obtain 
immediate feedback on their research work.

\subsection{Bibliometrics and usage analysis}
\label{usage}
\par For many reasons researchers, as authors and as readers, need 
the attribution of quality indicators to articles. Defining such 
'quality' has always been a challenge and the problem of evaluating 
archive content in an objective, unbiased, fashion is therefore not 
new. Traditionally, quality has been assigned simply by screening
through peer-review in combination with the measured impact of the 
journal, based on its referring component, i.e. the analysis of the 
scientific publications that cite it.  Although the procedure of 
collecting such bibliographic information was originally performed 
only to aid information retrieval, commercial \textit{ad hoc} 
services, most notably the ISI Science Citation Index \textregistered, 
have increasingly been used to determine the numbers of citations, 
and by implication the popularity and impact of journals, articles and 
authors. The availability of preprint and other author-disseminated 
versions of articles, sometimes without publication information 
attached, adds a new dimension to the problem of accurately measuring 
impact by citations. Conversely, the electronic era offers new 
possibilities for measuring access to, and usage of, an item. The 
definition of usage can therefore be expanded leading to new formulae 
for expressing an item's popularity and 'quality'.

\par In the digital era, such evaluation mechanisms have found
large-scale adoption, e.g. Google's PageRank that weights and ranks a
webpage based on an analysis of external hyperlinks that point to it,
and the University of Southampton's 
Citebase\footnote{http://www.citebase.org/} that maintains an open
citation index of OA on-line literature. At CERN, a similar
infrastructure is being worked on by extracting bibliographic
information from all publications in the repository (as explained in
section 4.1) and building a networked citation index. The advantages
of such an on-line open system, compared to traditional commercial
indexes are many: a) indexes are based on the whole collection present
in the archive, not a subset of the world's leading journals, b) the
availability of bibliographic information within a digital library's
metadata can be easily used to generate more complex reports and
ranking methods, e.g. co-citation, c) it is freely available, and 
authors are encouraged to self-archive their material to get a feel for 
their citation impact. 

\par The fact that full-text files of most physics publications are 
nowadays deposited in archives in digital formats, allows yet another 
method to evaluate the importance of research literature: usage. At 
CDS, usage analysis is performed by examining the server access logs 
in order to produce a) quantitative reports, such as ``most viewed'' or ``most
downloaded'' articles, and b) automatic recommendation reports, such as
``people who viewed this article also viewed''. This statistical
information, already largely adopted in non-scientific domains, is
rapidly acquiring consensus from the scientific community
\citep{pepelosa06}.

\section{Transformation of the publication landscape} 
\par The CDS service is a central part of many particle physicists' 
academic lives. Through enhanced searches and alerts, by clicking 
increasing numbers of relevant links, then by storing and sharing 
the resulting output, the communication of research results 
has been dramatically improved.

\par Although the physics preprint repository network offers free 
access to a very large proportion of the published material in the 
field, the community still relies on the publication system for 
validation of the final output. Rather than reinvent this system 
through the repositories, the community at CERN has expressed a 
preference for finding a way to transform publication into an OA 
system whereby the costs are moved from the readers to the authors 
who pay for the validation of their work. Repositories and OA 
publication, have always been seen by the OA community as 
complementary paths. With such a set-up it becomes possible to 
offer the finished articles to readers for no charge and to better 
integrate the articles into the repositories. The Report of the 
Task Force on Open Access Publishing in Particle Physics\citep{pepevoss06} 
and the potential new models that it proposes are currently in 
discussion.

\par For CERN, the start of the LHC era offers a perfect 
opportunity to launch into this new area of OA publication. 
The LHC -- Large Hadron Collider -- is the new accelerator 
being built at CERN and which is due to be completed in 2007. 
Once it is running, a new generation of physicists will begin 
their experiments and analysis, and along with the discoveries 
will come a large increase in the numbers of publications. It 
is hoped that the new physics can be accompanied by a new era 
of publishing -- one in which the results of the research are 
free to all and not limited only to those who continue to 
afford the journal subscriptions. Co-operation will continue 
with the traditional publishers in the field, and discussions 
should naturally lead on to new possibilities concerning Open 
Data, and perhaps repository overlay journals which could lead 
to quite radical changes in the way scientists prepare their 
publications. 

\par Whilst the OA movement is growing in recognition and 
acceptance, scholarly communication is under the microscope 
as never before, and interest in repositories is becoming an 
international library phenomenom. Increasing numbers of technical 
developers of repositories and thinkers behind the OA movement 
are emerging and generating new ideas. The CERN team has tried 
to realign its own direction with that of the worldwide movement 
and has also tried to keep a strong focus on innovation and 
experimentation. It is now an exciting time both for repository 
development and for the scholarly communication environment. CERN's 
position as a current leader in particle physics research, 
places it well, at least in its own field, for leading a change 
in the protocols for scholarly communication.

\acknowledgements The authors wish to thank the leaders of the 
Library and Document Server teams: Jens Vigen and Jean-Yves Le Meur, 
for encouraging participation in this conference.

\end{document}